\documentclass[11pt]{article}
\usepackage{hyperref}
\usepackage{authblk}
\usepackage{graphicx}
\usepackage{caption}
\usepackage{amsmath}
\usepackage{amssymb}
\usepackage{mathtools}
\usepackage{multirow}
\usepackage{amsmath}
\usepackage{color}
\usepackage{subcaption}
\usepackage{wasysym}
\usepackage{ulem}
\usepackage{hyperref}
\usepackage{authblk}
\usepackage{graphicx}
\usepackage{caption}
\usepackage{amsmath}
\usepackage{amssymb}
\usepackage{mathtools}
\usepackage{multirow}
\usepackage{amsmath}
\usepackage{color}
\usepackage{subcaption}
\usepackage{wasysym}
\usepackage{ulem}
\usepackage[utf8]{inputenc}
\usepackage[T1]{fontenc}
\usepackage{mathtools}
\usepackage[thinc]{esdiff}
\usepackage[bottom]{footmisc}
\usepackage{float}
\usepackage[euler]{textgreek}
\usepackage{siunitx}
\usepackage{graphicx}
\usepackage{dcolumn}
\usepackage{bm}
\usepackage{braket}
\usepackage{cite}
\usepackage{physics}
\usepackage{xfrac}
\usepackage{blindtext}
\usepackage{tikz}
\begin{document}

\title{Enhancement of Alpha Decay due to Medium Effects }

\author{Pankaj Jain$^{1}$\footnote{pkjain@iitk.ac.in}}
\author{Harishyam Kumar$^{2}$}
\affil{
$^1$ Department of Space, Planetary and Astronomical Sciences \& Engineering, Indian Institute of Technology, Kanpur, 208016, India\\
$^2$ Physics Department, Indian Institute of Technology, Kanpur, 208016, India
}

\maketitle

\begin{abstract}
We study the effect of medium on radioactive alpha decay and other similar
decays. The initial state in these type of decays is a quasi-bound state
with energy greater than zero. Such a state has very large amplitude in the
nuclear region and is exponentially suppressed at larger distances. The decay
rate of such states is known to decrease rapidly with decrease in the Q-value.
Here we study such a decay within a medium. We assume a simple spherically
symmetric repulsive potential model for the medium. This models the cumulative
effect of all nuclei in the medium which at short distances present
repulsive Coulomb interaction. We find that as the Q-value becomes very small,
the medium effects lead to a substantial enhancement in rate. In contrast,
for large Q-values, the medium effects are negligible. We briefly comment
on application to real systems and the experimental implications of this result.
\end{abstract}

\section{Introduction}

The decay rates of radioactive nuclei are known to 
show some dependence on environment \cite{Segre47,daudel1947alteration,PhysRev.90.430,PhysRev.112.77,PhysRev.117.795,ALDER1969487,ALDER1971163,doi:10.1126/science.181.4105.1164,LIU2000163,PhysRevLett.29.1188,JCMNS2018,PhysRevC.101.035801,Sikdar2022,ARayEPJC}. This is not too surprising since it 
is possible to make small changes in nuclear potential by changing the 
environment.   
It has been observed that 
the $\beta^-$ decay rate of some nuclei can change very strongly
if they are strongly ionized in comparison to being in neutral state
\cite{PhysRevLett.77.5190}. We point out that 
theoretical calculations do not lead to good agreement with some
of the observations \cite{ARayEPJC}. Furthermore, there exists unexplained
dependence of alpha decay rate under lattice compression \cite{Sikdar2022}.
In the present paper we are interested in the medium dependence of the 
alpha-decay process along with other similar processes. 
The initial state in this 
case is a quasi-bound state with energy eigenvalue $E>0$.
The radioactive $\alpha$-decay rate can be calculated by standard 
methods \cite{Krane:359790,PhysRev.113.1593,PhysRev.119.1069,PhysRev.169.818,SANDULESCU1978205,PhysRevLett.59.262,Holstein1996}.  
It is generally believed that such processes will have
relatively small medium dependence. 
For most cases, which
have been studied, the final state alpha particle has 
relatively high energy compared to the typical medium potential. 
For such cases the medium effects are likely to be
negligible. 

\begin{figure}[h]
	\centering
	\includegraphics[ clip,scale=0.9]{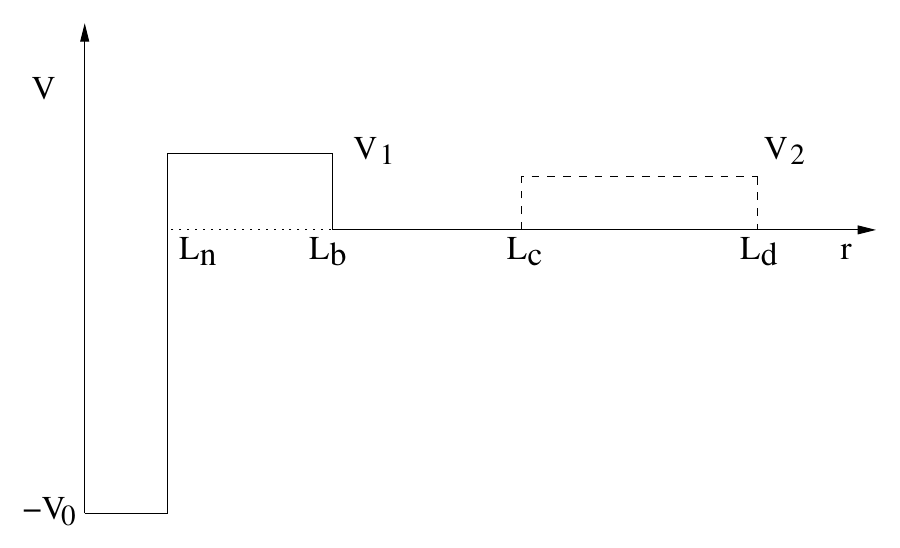}
\caption{\label{fig:potential} Schematic illustration of model potential}
\end{figure}

In the present paper we consider the case in which the $Q$ value of the process is relatively small,
leading to a small final state energy of the alpha particle. As we shall see, in this case 
the medium can affect such a process in a more dramatic manner.
In free space the relevant quasi-bound
state has a very fine
tuned wavefunction which decays exponentially through the barrier. Let us
consider this through a simple model in which we consider piece-wise constant
potentials. This is convenient and a more complex nuclear potential along
with Coulomb barrier only involves more mathematical complexity which can be
easily included once the basic mechanism is clear. 
Let us consider a spherically symmetric potential whose radial 
dependence is shown in Fig. \ref{fig:potential}. The solid
line shows the potential corresponding to a nuclear well along with a 
barrier and finally free space at large distances. 
We also assume an effective model for the medium potential which is shown in Fig. \ref{fig:potential}
by dashed line. 
This simply represents an average repulsive
effect of all the ions in the medium. We may also include attractive 
contribution of the electrons but that does not introduce any major change
in the argument.  

We assume that initially the nuclei are in free space and are introduced into the medium at some
time. Hence, the medium potential may be taken to be time dependent. It is zero initially and 
eventually takes its full value.
The time dependence of the potential is governed by the process by which the
nuclei are introduced into the condensed medium.  
Here we adopt a simple model by keeping the width of the medium potential
fixed and allowing its height
$V_2$ to be time dependent. Hence, we
 express the potential as,
\begin{equation}\label{eq:pot1}
	V(r,t) =
	\begin{cases}
		-V_0 & \text{$0\leq r <L_n$ }\\
		V_1 & \text{$L_n \leq r < L_b$}\\
		0 & \text{$L_b \leq r < L_c$}\\
		V_2(t) & \text{ $L_c \leq r < L_d$}\\
		0 & \text{$r \ge L_d$}
	\end{cases}    
\end{equation}
where $V_2(t=0)=0$ and $V_2(t\ge T) = V_{2f}$. Throughout this paper we will assume $V_2(t) < V_1$.  
We are interested in the wave function of a particle Y which experiences
the potential $V(r)$ due to a nucleus X along with the medium potential. The energy eigenvalue of this 
state is taken to be greater than zero. 
For the state under consideration, at $t=0$, the wave function would have an exponentially 
decaying dependence on $r$ in the range $L_n\le r < L_b$ and finally 
show free space dependence for larger $r$. 
We point out that the wave function displays this behaviour only over a 
very narrow range of energies centered at the energy $E=E_R$ 
and we refer to the corresponding state as
a resonant state. As we deviate from this energy, the amplitude becomes very
small at nuclear distances and large at larger distances.
The corresponding resonant or 
quasi-bound state eventually decays leading to a free X particle.
In free space, the decay rate is governed by 
the exponential suppression factor due to the repulsive barrier. 

Inside a medium the wave function can be dramatically modified by the medium potential,
represented in Fig. \ref{fig:potential} by $V_2(t)$.
The wave function for the quasi-bound state remains unchanged for $r<L_c$.
However, unless the medium potential is fine tuned at $r=L_c$, 
the wave function will develop an exponentially growing factor in the second
barrier, i.e. for $r> L_c$. The growth is controlled by the width and the
height of this barrier. Hence, the wave function in free space i.e. for $r>L_d$
may change significantly due to the presence of second barrier. In an 
extreme case, it could become larger than the wave function inside the 
nuclear well and hence completely change the nature of the quasi-bound
state. We expect this can have major implications for the decay rate for
this state, which we study in the present paper.

In application to real nuclei, we need to identify cases in which a parent nuclei can decay to 
two or more daughter nuclei with a relatively small $Q$ value. We assume that a medium provides a 
confining potential which can accelerate the decay process. The confining potential can be enhanced
by loading impurities into the system.  
Alternatively, we may directly impose an external repulsive potential.

\section{Basic Mechanism}

We are interested in the decay of a quasi-bound state of particle X of mass $m$ 
in the potential
of nucleus Y, which is taken to very massive.
The nucleus X is assumed to be initially bound to nucleus Y and eventually
becomes a free particle.
The potential model is given by Eqs. \ref{eq:pot1} 
 where $\vec r$ represents the relative coordinate.
Here we focus on the relative coordinate $\vec{r}$ since  
the centre of the mass coordinate does not play any essential role.
As described above, Eq. \ref{eq:pot1} represents the attractive
nuclear potential along with the repulsive potential at larger $r$, including medium effects.
A possible set of parameters are $V_0=50$ MeV, $V_1=100$ in atomic units (a.u.),
the nuclear length scale $L_n=0.9566 \times 10^{-4}$ a.u. and the 
barrier length scale $L_b=0.1$ a.u. \cite{Kumar2025}. The precise value
of $L_n$ is chosen for convenience in obtaining the nuclear bound state whose
parameters are given below. Furthermore, we set $L_c = 1$, $L_d=2.5$, $V_{2f}=1$ and time $T=1$.

Let us assume that the state under consideration has $l=0$. 
The instantaneous eigenfunctions corresponding to the potential given in Eq. \ref{eq:pot1} may be expressed as, 
\begin{equation}
	\psi_k(r,t)=  
	{1\over\sqrt{4 \pi}}
	\frac{U_k(r,t)}{r}
\label{eq:inswavefn}
\end{equation}
Here $k=\sqrt{2mE}$ represents the wave number and $E$ the energy eigenvalue. 
The wave functions $U_k(r,t)$ at any fixed time for different cases corresponding to $E\ge 0$ are given in Appendix A.
There also exist discrete states with $E<0$, corresponding to the nuclear bound states. The eigenfunctions for these states are 
very strongly suppressed beyond nuclear distances and expected to play negligible role in our analysis. Hence,
we ignore these in our discussion. 
We point out that all eigenfunctions for $E\ge 0$ behave as free particle states for $r>L_d$. Hence,
we normalize them as free particle states with the upper limit $L$ on $r$ taken to infinity at the
end of the calculation. 
The normalization constant for these states is given by
\begin{equation}
	N =  \sqrt{{L\over 2}\, (J^2+M^2)}
\end{equation}

We are interested in a state in which the nuclei Y and X initially form a quasi-bound state. This state is close
to the eigenstate for $V_2=0$ for which the coefficient $C$ in Eq. \ref{eq:coeffs} is zero. This is the
condition for resonance and we denote the corresponding energy by $E_R$. 
This state is a superposition of eigenstates and hence undergoes
time evolution.  
The wave function at any time for $l=0$ can be expressed as,
\begin{equation}
	\psi(r,t)=  
	{1\over\sqrt{4 \pi}}
	\frac{U(r,t)}{r}
\end{equation}
where $U(r,t)$ satisfies the radial Schrodinger equation,
\begin{equation}
	i\hbar {\partial U\over \partial t} = -{\hbar^2\over 2m}{\partial^2U\over\partial r^2} + V(r,t) U
\end{equation}
We consider the following wave-packet model for the initial state, corresponding to $t=0$, 
with 
\begin{equation}
U(r,0) = \int_0^\infty dk U_k(r,0)a_k \,.
	\label{eq:initialU}
\end{equation}
with 
\begin{equation}
a_k = {A\over (k-k_R)^2+\Gamma^2}\,.
\label{eq:wavepacket}
\end{equation}
Here $U_k(r,0)$ are the eigenstates for the potential with $V_2=0$, $k_R$
represents the corresponding resonant wave-number and $\Gamma$ is the width 
parameter, which is expected to be very small, of order $e^{-2\kappa_{2R}L_b}$. 
Here $\kappa_{2R}$ is the value of $\kappa_2$ at 
resonant energy. 
The normalization constant $A$ is fixed by the condition,
\begin{equation}
	\int_0^\infty dr U^*(r,0) U(r,0) =1 
\end{equation}
i.e. the wave function $U(r,0)$ is normalized to unity.
This condition leads to 
\begin{equation}
	A^2 = {2L\Gamma^3\over \pi^2}\,. 
\end{equation}
It is clear that we get significant
contributions only from states close to the resonant state. Furthermore,
as long as $V_2=0$,
we expect the lifetime of this state to be very large, determined by the factor 
$1/\Gamma$. The width $\Gamma$ can be calculated by standard methods 
\cite{Krane:359790,PhysRev.113.1593,PhysRev.119.1069,PhysRev.169.818,PhysRevLett.59.262,Holstein1996}.  

Another possible choice of the initial
wave function is the following, 
\begin{eqnarray}
	U_i(r) &=& {1\over N_i} \sin k_{1i}r\ \ \ \ \ \ \ \ r < L_n\nonumber \\ 
	 &=& {B_i\over N_i} e^{-\kappa_{2i}r} \ \ \ \ \ \ \ \ r \ge L_n\,, 
	\label{eq:initialU1}
\end{eqnarray}
\begin{equation}
	N_i^2= {1\over 2}\left[L_n-{\sin 2k_iL_n\over 2k_i}\right] + 
	B_i^2\, {e^{-2\kappa_{2i}L_n}\over 2\kappa_{2i}}\,, 
\end{equation}
$k_{1i}=\sqrt{2m(E_i+V_0)}$, $\kappa_{2i} = \sqrt{2m(V_1-E_i)}$ and $B_i$ is given by Eq. \ref{eq:coeffs} 
with $k_{1}$ and $\kappa_2$ replaced by $k_{1i}$ and $\kappa_{2i}$ respectively.
Here $E_i=E_R$ where $E_R$ is the resonant energy, as defined above. Hence, 
this state is same
as the resonant state up to $r= L_b$ but beyond that it decays rapidly. 
This state, however, has relatively large overlap with eigenstates with 
energies $E>V_1$. This results in a fairly large decay rate of this state
even in free space. The decay rate expected to be controlled by the 
exponential suppression factor $e^{-2\kappa_{2R}L_b}$. However, the 
contribution from states with $E>V_1$ is not suppressed by this factor and
leads to a decay rate orders of magnitude larger than the expected rate
for the resonant energy under consideration.
Hence, this state 
does not properly represent the experimental conditions, and we choose
our initial state as given in Eq. \ref{eq:initialU}. 

For the case $V_2=0$, the initial state given in Eq. \ref{eq:initialU} will slowly evolve over a time
scale governed by the exponential factor $e^{2\kappa_2L_b}$, such that it's 
amplitude inside the nucleus decays and builds outside the nuclear region.
Hence, this effectively leads to a slow decay of the quasi-bound state.
However, let us consider the situation
such that after some time $V_{2}$ is sufficiently large and leads to 
\begin{equation}
	e^{2\kappa_3(L_d-L_c)} \sim e^{2\kappa_2L_b}
\end{equation}
In this case the resonant wave function ($E=E_R$) at $r<L_n$ would be 
comparable to that at $r>L_d$. Hence, the integrated amplitude for $r<L_n$ would
be much smaller than that at larger $r$ since $L_n<< L_d$. The eigenfunctions
for all other values of $E$ would also be strongly suppressed at $r<L_n$. Hence we expect that beyond this
time the initial state would decay quickly in the nuclear region. We point out that if $e^{2\kappa_3(L_d-L_c)}$ is
much larger than $ e^{2\kappa_2L_b} $, the amplitude at nuclear distances would
be very small in comparison to that at larger distances.
We develop the methodology to determine the time evolution of this state in the
next section.

\section{Evolution of the wave function}
The wave function $U(r,t)$ at any time $t$ can be expressed as, 
\begin{equation}
U(r,t) = \int_0^\infty dk e^{-iEt/ \hbar} a_k(t) U_k(r,t)
	\label{eq:Urt}
\end{equation}
where $U_k(r,t)$ are related to the instantaneous eigenstates $\psi_k(r,t)$ of the Hamiltonian
by Eq. \ref{eq:inswavefn}. The latter satisfy,
\begin{equation}
	H(t)\psi_k(r,t) = E \psi_k(r,t) 
\end{equation}
corresponding to the energy eigenvalue $E$.
Here $E$ is independent of time and each state is labelled by $E=\hbar^2k^2/2m$. We point out that
we have additional discrete energy eigenvalues $E<0$ which correspond to nuclear bound states and are
being ignored in this analysis. Only the continuum $E>0$ states which give dominant contribution are included.
By substituting this form of the wave function in the Schrodinger equation, we obtain
\begin{equation}
\dot a_{k'}(t) = -{L\over \pi}\int_0^\infty {dk} a_{k}(t) e^{-i(E-E')t/\hbar} \int_0^\infty dr U_{k'}^*\dot U_{k} 
	\label{eq:evolution}
\end{equation}
where $\hbar k' = \sqrt{2mE'}$\,.
The 
initial condition to be imposed on $a_k$ is given by Eq. \ref{eq:wavepacket},
\begin{equation}
	a_k(0) = {A\over (k-k_R)^2+\Gamma^2}\,.
\label{eq:initial}
\end{equation}
Solution to this integral equation (Eq. \ref{eq:evolution})
provides the time evolution of the
coefficients $a_k(t)$ and hence the complete wave function $U(r,t)$. 

\subsection{Solution to the evolution equation}
The intantaneous eigenfunctions for the potential given in Eq. \ref{eq:pot1}
are given in Appendix A. 
We first determine the radial integral,
\begin{equation}
I = \int_0^\infty dr U_{k'}^*\dot U_{k}
\end{equation}
which appears in Eq. \ref{eq:evolution}. We express $U_k= f_k/N_k$ and obtain
\begin{equation}
\dot U_k = -{\dot N_k\over N_k} U_k + {1\over N_k}\dot f_k
\end{equation}
This leads to, 
\begin{equation}
I = -{\dot N_k\over N_k} {\pi\over L}\delta(k-k') + {1\over N_k}
	\int_0^\infty dr U^*_{k'} \dot f_k
	\label{eq:integralI}
\end{equation}
We may express the second term on the right hand side as,
\begin{equation}
 {1\over N_k}
	\int_0^\infty U^*_{k'} \dot f_k
= I_1 + I_2
\end{equation}
where $I_1$ and $I_2$ get contributions from the range $L_c\le r < L_d$ and 
$r\ge L_d$ respectively. 

For $E>V_2$ and $E'>V_2$, we obtain
\begin{eqnarray}
	I_1 &=& {1\over N_kN_{k'}}\int_{L_c}^{L_d} dr \left[G(E')\sin k'_3r
+ H(E')\cos k'_3 r\right] \nonumber\\
	&\times& \left[\dot G(E)\sin k_3r+\dot H(E)\cos k_3r + 
	G(E)\dot k_3r\cos k_3r - H\dot k_3r\sin k_3r \right]
\end{eqnarray}
and 
\begin{equation}
	I_2 = {1\over N_kN_{k'}}\int_{L_d}^{\infty} dr \left[J(E')\sin k'r
	+M(E')\cos k'r\right]\left[\dot J(E)\sin kr + \dot M(E) \cos kr\right]
\end{equation}
For $E<V_2$ and $E'>V_2$, $I_2$ has the same form, while $I_1$ is given by,
\begin{eqnarray}
I_1 &=& {1\over N_kN_{k'}}\int_{L_c}^{L_d} dr \left[G(E')\sin k'_3r
+ H(E')\cos k'_3 r\right] \nonumber\\
	&\times& \Bigg\lbrace e^{-(r-L_c)\kappa_3}\left[ {d {\bar G(E)}\over dt}-
	\bar G(r-L_c)\dot\kappa_3\right] + e^{(r-L_c)\kappa_3} 
	\left[{d {\bar H(E)}\over dt}  + 
	\bar H(E) (r-L_c)\dot \kappa_3 \right]\Bigg\rbrace
\end{eqnarray}
Here we have defined $\bar G$ and $\bar H$, such that $G = e^{\kappa_3L_c}\bar G$ and $H = e^{-\kappa_3L_c}\bar H$.
Similarly for the remaining two cases, corresponding to $E'<V_2$, $I_2$ takes
the same form, while we need to replace $U_{k'}$ with the wave function 
corresponding to $E'<V_2$.

The integral $I_1$ can be evaluated directly for all the cases, 
while the integral $I_2$ can
be expressed as a sum of four terms $I_{21}$, $I_{22}$, $I_{23}$ and $I_{24}$.
We obtain,
\begin{eqnarray}
	I_{21} &=& {1\over N_kN_{k'}}\int_{L_d}^{\infty} dr J(E')\dot J(E)\sin k'r\sin kr\nonumber\\
	&=& {J(E')\dot J(E)\over N_kN_{k'}}\left[{\pi\over 2}\delta(k'-k)
	- \int_0^{L_d} dr \sin k'r\sin kr\right]
	\label{eq:I21}
\end{eqnarray}
Here we have expressed the integral as difference of integrals from $0$ to 
$\infty$ and from $0$ to $L_d$ and used the orthogonality of $\sin kr$ and 
$\sin k'r$. We next evaluate,
\begin{equation}
	I_{22} = {J(E')\dot M(E)\over N_kN_{k'}}  \int_{L_d}^\infty
dr	\sin k'r\cos kr
\end{equation}
In this case we cannot use orthogonality and hence directly evaluate the 
integral by taking the upper limit as $L$ which would eventually be taken to
infinity. We obtain,
\begin{equation}
	I_{22} = {J(E')\dot M(E)\over N_kN_{k'}}{1\over 2} 
	\left[-{\cos(k'+k)L\over k'+k} -{\cos(k'-k)L\over k'-k}
	+ {\cos(k'+k)L_d\over k'+k} + {\cos(k'-k)L_d\over k'-k}\right]
\end{equation}
The first term on the right hand side oscillates very rapidly with $k$ (or $E$)
in the limit $L\rightarrow \infty$. Hence, it will give negligible contribution
to the integral over $k$ in Eq. \ref{eq:evolution} and can be dropped. The only
potential problem is in the case $k'=0$. In this case there may be a divergence
in the limit $k\rightarrow 0$. However, after integration over $k$, we 
obtain negligible contribution from the lower limit of the integral over $k$.
This also applies to all the terms on the right hand side. The second term
on the right hand side also oscillates very rapidly with $k$ and gives 
negiligible contribution to the energy integral. In this case there is a 
potential divergence in the limit $k\rightarrow k'$. However, for any value of $L$, this
divergence cancels with the corresponding divergence in the last term. 
Hence, we do not expect the 
integral over $k$ in Eq. \ref{eq:evolution} to blow up.
We can essentially drop the first two terms in 
$I_{22}$ with the understanding that the divergence in the last term is cancelled 
by the second term. 
The same arguments apply to the integral $I_{23}$, given by,
\begin{equation}
	I_{23} = {M(E')\dot J(E)\over N_kN_{k'}}{1\over 2} 
	\left[-{\cos(k+k')L\over k+k'} -{\cos(k-k')L\over k-k'}
	+ {\cos(k+k')L_d\over k+k'} + {\cos(k-k')L_d\over k-k'}\right] \,.
\end{equation}
Finally, the integral $I_{24}$ is given by,
\begin{eqnarray}
	I_{24} &=& {M(E')\dot M(E)\over N_kN_{k'}}\int_{L_d}^{\infty} dr 
\cos k'r\cos kr\nonumber\\
	&=& {M(E')\dot M(E)\over N_kN_{k'}}\left[{\pi\over 2}\delta(k'-k)
	- \int_0^{L_d} dr \cos k'r\cos kr\right]
	\label{eq:I24}
\end{eqnarray}

The delta function terms in $I_{21}$ and $I_{24}$ essentially cancel the
first term on the right hand side of Eq. \ref{eq:integralI}. Hence only
$I_1$ and the remaining terms in $I_2$ contribute to $I$. We denote these collectively
as $\Delta I$, i.e.,
\begin{equation}
	\Delta I = I_1 + \tilde I_2
\end{equation}
where $\tilde I_2$ is equal to $I_2$ without the delta function terms in $I_{21}$
(Eq. \ref{eq:I21}) and $I_{24}$ (Eq. \ref{eq:I24}). 
Hence the evolution equation can be expressed as,
\begin{equation}
\dot a_{k'}(t) = -{L\over \pi}\int_0^\infty {dk} a_{k}(t) e^{-i(E-E')t/\hbar} \Delta I\,. 
	\label{eq:evolution1}
\end{equation}
We seek a solution of the form
\begin{equation}
	a_k(t) = {A\over (k-k_R)^2+\Gamma^2} +\delta a_k(t)
	\label{eq:aEt}
\end{equation}
where the first term on the right hand side is equal to $a_k(0)$. Substituting this
into Eq. \ref{eq:evolution1}, we obtain,
\begin{equation}
{\partial \delta a_{k'}(t)\over \partial t} = 	 	
-{L\over \pi }\int_0^\infty {dk} {Ae^{-i(E-E')t/\hbar}\over (k-k_R)^2+\Gamma^2}  \Delta I
-{L\over \pi }\int_0^\infty {dk} \delta a_{k}(t) e^{-i(E-E')t/\hbar} \Delta I\,.  
\label{eq:evolution2}\end{equation}
This equation can be solved numerically. The second integral on the right hand side can
be evaluated directly, while the first is best evaluated by making a change of variables.

As explained earlier, $\Gamma$ is of order $e^{-2\kappa_{2R}L_b}$ and, hence,
very small. We get dominant contributions to the first integral on the right hand side of 
Eq. \ref{eq:evolution2}
only for $k$ very close to $k_R$. Hence, in order to evaluate this integral
we define the parameter $x$, given by,
\begin{equation}
	x={k-k_R\over \Gamma} 
	\label{eq:parameterx}
\end{equation}
which leads to $dk = \Gamma dx$. 
The limits on $x$ range from $-X$ to $X$ where $X$ is taken to be a large number, such as, 1000. The integral is not expected to be too sensitive to its precise value, assuming that it is sufficiently large. 
This is because the integrand decays rapidly with $x$.
All the terms which have a mild dependence 
on $k$ near resonance can be approximated by their value at resonance. 
The only term which shows very rapid dependence is the term corresponding
to the coefficient $C_k$. It is useful to express this as
\begin{equation}
	C_k= (E-E_R) \bar C
\end{equation}
where
\begin{equation}
        \bar C = {e^{-\kappa_{2R}L_n} \sqrt{m_p}\over 2\sqrt{2}\hbar \kappa_{2R}}
        \left({C_1\over \sqrt{V_0+E_R}} - {\sin k_{1R}L_n\over \sqrt{V_1-E_R}}
        \right)
\end{equation}
and
\begin{equation}
        C_1 = \kappa_{2R} L_n\cos k_{1R}L_n
        -k_{1R}L_n\sin k_{1R}L_n + \cos k_{1R}L_n
\end{equation}
Here $k_{1R}$ and $B_R$ refer to the values of 
$k_1$ and $B_n$ respectively at the energy $E=E_R$.
Replacing $E$ in terms of $k$, we obtain
\begin{equation}
	C_k = {\bar Ck_R\over m}x\Gamma 
\end{equation}
where we have approximated $k+k_R\approx 2k_R$. 
Furthermore we set
\begin{equation}
\Gamma = \beta e^{-2\kappa_{2R}L_b} 
\end{equation}
The pre-factor $\beta$ can be estimated by standard methods
\cite{Krane:359790,PhysRev.113.1593,PhysRev.119.1069,PhysRev.169.818,PhysRevLett.59.262,Holstein1996}.  
Hence, we obtain
\begin{equation}
	C_ke^{\kappa_{2R}L_b} = {\bar Ck_R\over m}x\beta e^{-\kappa_{2R}L_b} 
\end{equation}
Substituting this into the wave function, we find that it is useful to take
out an overall factor $e^{-\kappa_{2R}L_b}$ from the coefficients $D_k$, $F_k$,
$G_k$, $H_k$, $J_k$ and $M_k$. Hence we replace these in terms of 
$\tilde D_ke^{-\kappa_{2R}L_b}$,
$\tilde F_ke^{-\kappa_{2R}L_b}$,
$\tilde G_ke^{-\kappa_{2R}L_b}$,
$\tilde H_ke^{-\kappa_{2R}L_b}$,
$\tilde J_ke^{-\kappa_{2R}L_b}$ and 
$\tilde M_ke^{-\kappa_{2R}L_b}$
respectively.
The normalization factor can be written as
\begin{equation}
	N_k = \sqrt{L/2} e^{-\kappa_{2R}L_b}\sqrt{\tilde J^2 + \tilde M^2} 
\end{equation}


It is clear from Eq. \ref{eq:evolution2} that the correction factor $\delta a_k$ is of order $\sqrt{\Gamma}$
and hence suppressed for all values of $k$. This is determined by the 
inhomogeneous term on the right hand side of this equation. Hence, we find 
that for $k-k_R$ of order
$\Gamma$, the leading term in $a_k(t)$ (Eq. \ref{eq:aEt}) is of order $1/\sqrt{\Gamma}$ and hence
much larger than $\delta a_k$. For larger values of $k-k_R$, the leading term 
is of order $\Gamma^{3/2}$. In this case the correction factor, 
which is of order $\sqrt{\Gamma}$, dominates
over the leading term. In evaluating the time dependent wave function,
the contribution from this region is comparable to that from the small 
region $k-k_R$ of order
$\Gamma$. Hence, we cannot ignore this contribution, even though it 
appears suppressed. 
We point out that the factors of $L$ cancel out in the computation of 
$U(r,t)$. 
We can solve the evolution equation, Eq. \ref{eq:evolution1} numerically.

The wave function $U(r,t)$ can now be evaluated using Eq. \ref{eq:Urt} with
$a_k(t)$ given by Eq. \ref{eq:aEt}. The resulting wave function can be 
expressed as,
\begin{equation}
	U(r,t) = U_0(r,t) +\delta U(r,t)\,.
\end{equation}
Here,
 $U_0(r,t)$ is obtained from the first term on the right hand side of
 Eq. \ref{eq:aEt} and is given by,
\begin{equation}
U_0(r,t) = {\sqrt{2L\Gamma^3}\over \pi}\int_0^\infty dk e^{-iEt/ \hbar} {1\over (k-k_R)^2+\Gamma^2} U_k(r,t)\,.
	\label{eq:Urt1}
\end{equation}
and 
\begin{equation}
\delta U(r,t) = \int_0^\infty dk e^{-iEt/ \hbar}\delta a_k(t) U_k(r,t)
	\label{eq:Urt}
\end{equation}
In order to evaluate $U_0(r,t)$ we use the variable $x$, defined in Eq. 
\ref{eq:parameterx}. 
It is useful to express the eigenfunctions $U_k(r,t)$ as
\begin{equation}
	U_k(r,t) = {f_k(r,t) \over N_k}
\end{equation}
Using this we can express the wave functions as,
\begin{equation}
U_0(r,t) = {2\over \pi}\sqrt{\beta}e^{-iE_Rt/ \hbar}\int_{-X}^X   
	{dx\over x^2+1} {f_k(r,t)\over \sqrt{\tilde J^2 + \tilde M^2}} 
	\label{eq:Urt2}
\end{equation}

\section{Results}
We next numerically perform the integral in Eq. \ref{eq:Urt2} in order to determine $U_0(r,t)$. Furthermore, we solve the evolution equation, Eq. \ref{eq:evolution2} numerically to find the $\delta a_k(t)$ and compute the resulting 
contribution to the wave function $\delta U(r,t)$. The first integral over
$k$ on the right hand side of Eq. \ref{eq:evolution2} is computed by
using the variable $x$, defined in Eq. \ref{eq:parameterx}. The second
integral is evaluated numerically by changing the integration variable to
energy $E$ and imposing an upper cutoff of 12. We expect relatively small 
contribution from higher energy states.
The parameter $\beta$ is expected to be of order unity and we 
set it equal to 1. A more refined analysis can evaluate this more precisely, 
but the final result has some model dependence. This does not have affect our
conclusions and hence, we simply set this parameter to unity. 
We expect that for $t=0$, $U(r,0)$ will be of order unity for 
$r<L_n$ and very small for $r>L_b$. As time evolves and the potential $V_2$ 
grows, we expect, $U(r,t)$ will decay for $r<L_n$ and show relatively small
change for $r>L_b$. This behaviour is confirmed by our numerical calculation
for $\beta = 1$. 

\begin{figure}[h]
	\centering
	\includegraphics[ clip,scale=1.0]{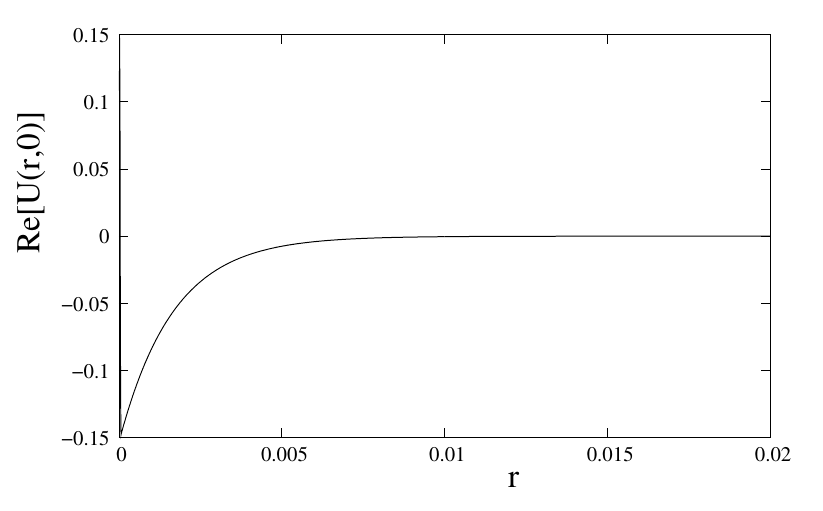}
	\caption{\label{fig:time0} The real part of the wave function $U(r,t)$ at $t=0$ for the chosen parameters. The imaginary part is zero at this time.}
\end{figure}

\begin{figure}[h]
	\centering
	\includegraphics[ clip,scale=1.0]{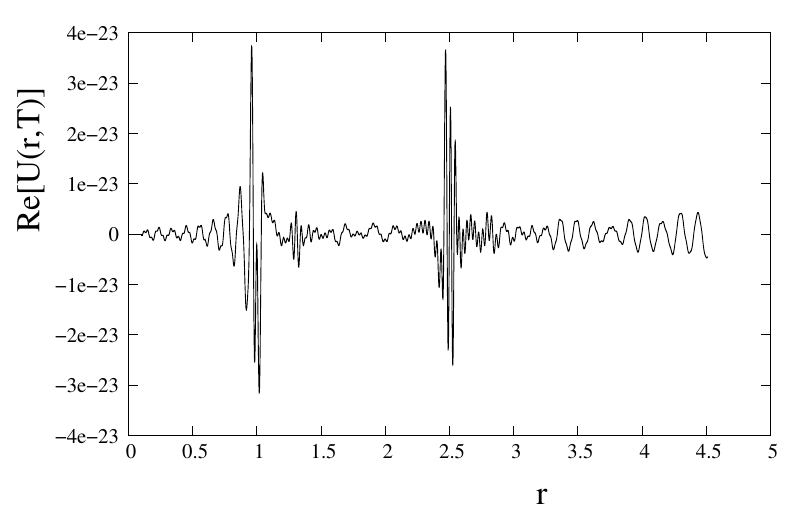}
\caption{\label{fig:timeT} The real part of the wave function $U(r,t)$ at $t=T$ for the chosen parameters. Here the time $T$ has been chosen to be unity. The imaginary part displays a similar behaviour.}
\end{figure}

In Fig. \ref{fig:time0} we show the real part of wave function $U(r,t)$ at
time $t=0$. At this time the potential $V_2=0$. The imaginary part is zero at this time. We find that the wave function is relatively large at nuclear distances and undergoes exponential decay 
at larger distances. It is very small of order $10^{-26}$ at $r>L_b$. This is
dictated by the factor $e^{-\kappa_2L_b}$. In Fig. \ref{fig:timeT}, we show the
real part of the wave function at $t=T$, where we have chosen the final time
$T=1$. At this time the potential $V_2$ takes its final value equal to 1. 
The wave function is of order $10^{-26}$ at nuclear distances and takes a 
somewhat larger value at
larger distance $r>L_b$. The imaginary part 
of the wave function displays a similar behaviour. We point out that the precise value is dictated by our choice of parameters which lead to the factor 
$e^{\kappa_3(L_d-L_c)}$ comparable to $e^{\kappa_2L_b}$. The wave function is very small outside the nuclear region at both times, but extends to very large distances and is normalized to unity at both times. We find that at $t=T$, the probability to find the particle at nuclear distances has reduced drastically. This can be further reduced by chosing a larger value of $L_d$ or $V_2$. This decay has taken place in the time scale chosen to impose the external potential and is very small compared to the natural decay time of this state.  Hence, we find that in such special cases, the medium can have a very strong effect on the decay of these quasi-bound states.

\subsection{Applications}
The proposed mechanism is likely to have many practical applications. 
Essentially, it leads to the conclusion that we
can enhance the decay rate of a nucleus by applying external
potential. Hence, if the energy of the quasi-bound state is relatively
small, of the order of tens of keV or smaller, it is possible to induce
its decay by applying repulsive electric potential which is relatively
easy to create in a laboratory. Besides this, the medium nuclei also  
exert a repulsive potential at short distances. The combined effect of 
many such nuclei will lead to a repulsive potential on the nucleus under
consideration in almost all directions. Such a potential will act as a cage and may be approximated roughly as the spherically symmetric potential we have used for our analysis. 
Hence, the decay rate of such nuclei may also be enhanced in a medium. 
If the energy of the quasi-bound state is very small, then it is likely to decay very rapidly in such a condensed medium and such nuclei may not have survived till today. However, if the energy is somewhat large, then it may survive. In this case, it may be possible to enhance the medium repulsive potential by inserting additional particles or impurities into the lattice. Such enhancement can accelerate the decay of such states. The values of energies for which this is applicable requires a detailed analysis of medium effects which we postpone to future research. Besides medium effects, it may also be possible to directly apply suitable external potential to meet the conditions to accelerate the decay rate. 

Our proposed mechanism might explain some some unusual nuclear phenomenon
seen experimentally. In particular, in an electrolysis experiment using nickel and palladium thin films \cite{Miley1996NUCLEARTI,SRINIVASAN2020233}, a large number of elements have been found both with atomic number less than and greater than that of the 
element present initially. Our proposal may explain the presence of elements
with atomic number less than the parent nuclei. A detailed investigation   
 with realistic nuclear potentials along with the screened Coulomb barrier and a model of the medium potential is needed for such applications. Elements with higher atomic number might arise due to the mechanisms proposed in \cite{Kumar2025,ramkumar2024low}. Furthermore, our proposal may explain the observed dependence of alpha decay on lattice compression \cite{Sikdar2022}.  

\section{Conclusions}
We have shown that if the Q-value of the $\alpha$-decay process is very small, then it's decay rate can be strongly influenced by an external repulsive potential. This electric  
potential may be applied directly on the nuclei. Alternatively if the nuclei
are present inside a condensed medium, the combined effect of the medium ions
may result in such a potential.   
Essentially the eigenfunction corresponding to the quasi-bound state gets very strongly distorted under such conditions and picks up a very large amplitude outside the nuclear region. 
Hence, the behaviour of this eigenstate changes drastically and leads to a 
large amplitude 
for the alpha 
particle to leak out of the nucleus, even if it
is initially confined inside the nucleus.
Once it is outside, it can be detected by an external probe, essentially leading
to a decay of the state. Our analysis is also applicable to processes in which the emitted particle may not be the alpha particle but some heavier particle, such as, Carbon or Oxygen. Although such decays are not seen in nature, we cannot rule out the possibility that these are possible but have very small decay rates which are not accessible experimentally. In these cases also, the decay can be accelerated in processes for which the Q-value is sufficiently low. Our results have many applications in nuclear physics and nuclear astrophysics which can be probed in future.  

\bigskip
\noindent
{\bf Acknowledgements:} PJ is grateful to John P. Ralston for very useful 
discussions.

\section{Appendix A}\label{sec:appA}
In this appendix, we give the explicit form of the instantaneous wave functions $U(r)$ for different cases.
For the case, $E<V_2$, the wave function is given by, 
\begin{equation}
	U(r) =
	\begin{cases}
\frac{1}{N_k} \sin{k_1 r} & \text{$0\leq r < L_n$}\\
\frac{1}{N_k} [ B_k e^{-\kappa_2 r}+C_k e^{\kappa_2 r}] & \text{$L_n\leq r <L_b $}\\
\frac{1}{N_k} [ D_k \sin{k r }+F_k \cos{k r }] & \text{$L_b \leq r < L_c$}\\
		\frac{1}{N_k}[{G_k} e^{-\kappa_3 r} + H_k e^{\kappa_3r} ]& \text{$L_c \leq r < L_d$}\\
	\frac{1}{N_k}[{J_k} \sin{kr} + M_k \cos{kr} ]& \text{ $r \geq L_d$}
	\end{cases}       
\end{equation}
where,
\begin{equation} 
	\begin{split}
		\hbar k_1 & =\sqrt{2 m(V_0+E)} \\
		\hbar \kappa_2 & =\sqrt{2 m(V_1-E)}\\
		\hbar k &=\sqrt{2 m E} \\
		\hbar\kappa_3 &=\sqrt{2 m(V_2-E)}
	\end{split}
	\label{eq2:k_values}
\end{equation}
Here $N_k$ is the normalization constant.

 The coefficients are given by, 
\begin{eqnarray} \label{eq:coeffs}
	B_k & =& {1\over 2}e^{\kappa_2 L_n} \left(\sin k_1 L_n-{k_1\over \kappa_2} \cos k_1 L_n\right) \nonumber\\
	C_k & =& {1\over 2}e^{-\kappa_2 L_n} \left(\sin k_1 L_n+{k_1\over \kappa_2} \cos k_1 L_n\right)\nonumber\\
	D_k &=&	(B_k e^{-\kappa_2 L_b}+ C_k e^{\kappa_2 L_b}) \sin k L_b -
\frac{\kappa_2}{k}(B_k e^{-\kappa_2 L_b}- C_k e^{\kappa_2 L_b}) \cos k L_b\nonumber\\
	F_k &=&	(B_k e^{-\kappa_2 L_b}+ C_k e^{\kappa_2 L_b}) \cos k L_b +
\frac{\kappa_2}{k}(B_k e^{-\kappa_2 L_b}- C_k e^{\kappa_2 L_b}) \sin k L_b
\end{eqnarray}
\begin{eqnarray}
G_k &=& {1\over 2}e^{\kappa_3 L_c}\left[ D_k \left(\sin k L_c- 
{k\over\kappa_3} \cos k L_c\right) + F_k\left(\cos kL_c
+ {k\over\kappa_3}\sin kL_c\right)\right] \nonumber \\
H_k &=&{1\over 2}e^{-\kappa_3 L_c}\left[ D_k \left(\sin k L_c+ 
{k\over\kappa_3} \cos k L_c\right) + F_k\left(\cos kL_c
- {k\over\kappa_3}\sin kL_c\right)\right] \nonumber \\
J_k & =&\sin kL_d\left(G_k e^{-\kappa_3L_d} + H_k e^{\kappa_3L_d}\right) 
+ {\kappa_3\over k}\cos kL_d\left(-G_ke^{-\kappa_3L_d} + H_ke^{\kappa_3L_d}\right)
\nonumber\\
M_k & =&\cos kL_d\left(G_ke^{-\kappa_3L_d} + H_ke^{\kappa_3L_d}\right) 
- {\kappa_3\over k}\sin kL_d\left(-G_ke^{-\kappa_3L_d} + H_ke^{\kappa_3L_d}\right)
\end{eqnarray}

For the case, $E>V_2$, the eigenfunction is same for $r<L_c$. For $r\ge L_c$ 
we obtain,
\begin{equation}
	U(r) =
	\begin{cases}
		\frac{1}{N_k}[{G_k} \sin k_3r + H_k \cos k_3r ]& \text{$L_c \leq r < L_d$}\\
	\frac{1}{N_k}[{J_k} \sin{kr} + M_k \cos{kr} ]& \text{ $r \geq L_d$}
	\end{cases}       
\end{equation}
where
\begin{equation}
	\hbar k_3 =\sqrt{2 m(E-V_2)}
\end{equation}
and
\begin{eqnarray}
G_k &=& \sin k_3 L_c\left[ D_k \sin k L_c+ 
F_k \cos k L_c\right] + \cos k_3L_c
	\left[D_k\cos kL_c- F_k\sin kL_c\right]{k\over k_3} \nonumber \\
H_k &=& \cos k_3 L_c\left[ D_k \sin k L_c+ 
F_k \cos k L_c\right] - \sin k_3L_c
	\left[D_k\cos kL_c- F_k\sin kL_c\right]{k\over k_3} \nonumber \\
	J_k &=& \sin kL_d\left[G_k\sin k_3L_d +H_k\cos k_3L_d\right]
	+ \cos kL_d \left[G_k\cos k_3L_d -H_k\sin k_3L_d\right]{k_3\over k}\nonumber\\
	M_k &=& \cos kL_d\left[G_k\sin k_3L_d +H_k\cos k_3L_d\right]
	- \sin kL_d \left[G_k\cos k_3L_d -H_k\sin k_3L_d\right]{k_3\over k}
	\end{eqnarray}

\bibliographystyle{ieeetr}
\bibliography{nuclear}

\begin{thebibliography}{10}

\bibitem{Segre47}
E.~Segr\`e, ``Possibility of altering the decay rate of a radio- active
  substance,'' {\em Phys. Rev.}, vol.~71, no.~4, p.~274, 1947.

\bibitem{daudel1947alteration}
R.~Daudel, ``Alteration of radioactive periods of the elements with the aid of
  chemical methods,'' {\em Rev. Sci}, vol.~85, p.~162, 1947.

\bibitem{PhysRev.90.430}
K.~T. Bainbridge, M.~Goldhaber, and E.~Wilson, ``Influence of the chemical
  state on the lifetime of a nuclear isomer, ${\mathrm{tc}}^{99m}$,'' {\em
  Phys. Rev.}, vol.~90, pp.~430--439, May 1953.

\bibitem{PhysRev.112.77}
D.~H. Byers and R.~Stump, ``Low-temperature influence on the technetium-$99m$
  lifetime,'' {\em Phys. Rev.}, vol.~112, pp.~77--79, Oct 1958.

\bibitem{PhysRev.117.795}
R.~A. Porter and W.~G. McMillan, ``Effect of compression on the decay rate of
  ${\mathrm{tc}}^{99m}$ metal,'' {\em Phys. Rev.}, vol.~117, pp.~795--800, Feb
  1960.

\bibitem{ALDER1969487}
K.~Alder, J.~Hadermann, and U.~Raff, ``Environmental effects on the electron
  capture decay rate,'' {\em Physics Letters A}, vol.~30, no.~9, pp.~487--488,
  1969.

\bibitem{ALDER1971163}
K.~Alder, G.~Baur, and U.~Raff, ``Influence of the chemical environement on
  α-decay,'' {\em Physics Letters A}, vol.~34, no.~3, pp.~163--164, 1971.

\bibitem{doi:10.1126/science.181.4105.1164}
W.~K. Hensley, W.~A. Bassett, and J.~R. Huizenga, ``Pressure dependence of the
  radioactive decay constant of beryllium-7,'' {\em Science}, vol.~181,
  no.~4105, pp.~1164--1165, 1973.

\bibitem{LIU2000163}
L.~gun Liu and C.-A. Huh, ``Effect of pressure on the decay rate of 7be,'' {\em
  Earth and Planetary Science Letters}, vol.~180, no.~1, pp.~163--167, 2000.

\bibitem{PhysRevLett.29.1188}
M.~N. de~Mevergnies, ``Perturbation of the $^{235m}\mathrm{U}$ decay rate by
  implantation in transition metals,'' {\em Phys. Rev. Lett.}, vol.~29,
  pp.~1188--1191, Oct 1972.

\bibitem{JCMNS2018}
F.~Metzler, P.~Hagelstein, and S.~Lu, ``Observation of non-exponential decay in
  x-ray and γ emission lines from co-57,'' {\em Journal of Condensed Matter
  Nuclear Science}, vol.~27, 11 2018.

\bibitem{PhysRevC.101.035801}
A.~Ray, A.~K. Sikdar, P.~Das, S.~Pathak, and J.~Datta, ``Unexpected increase of
  $^{7}\mathrm{Be}$ decay rate under compression,'' {\em Phys. Rev. C},
  vol.~101, p.~035801, Mar 2020.

\bibitem{Sikdar2022}
A.~Sikdar, J.~Nandi, J.~Patel, D.~Pandit, J.~Datta, P.~Das, R.~Baidya, and
  A.~Ray, ``Does the alpha decay rate change under lattice compression?,'' 12
  2022.

\bibitem{ARayEPJC}
A.~Ray, P.~Das, A.~Sikdar, S.~Pathak, N.~Aquino, M.~Lozano-Espinosa, and
  A.~Artemyev, ``Electron capture nuclear decay rate under compression in a
  confined environment,'' {\em The European Physical Journal D}, vol.~75,
  p.~140, 04 2021.

\bibitem{PhysRevLett.77.5190}
F.~Bosch, T.~Faestermann, J.~Friese, F.~Heine, P.~Kienle, E.~Wefers,
  K.~Zeitelhack, K.~Beckert, B.~Franzke, O.~Klepper, C.~Kozhuharov, G.~Menzel,
  R.~Moshammer, F.~Nolden, H.~Reich, B.~Schlitt, M.~Steck, T.~St\"ohlker,
  T.~Winkler, and K.~Takahashi, ``Observation of bound-state
  ${\mathit{\ensuremath{\beta}}}^{\ensuremath{-}}$ decay of fully ionized
  ${}^{187}$re: ${}^{187}$re${\ensuremath{-}}^{187}$os cosmochronometry,'' {\em
  Phys. Rev. Lett.}, vol.~77, pp.~5190--5193, Dec 1996.

\bibitem{Krane:359790}
K.~S. Krane, {\em {Introductory nuclear physics}}.
\newblock New York, NY: Wiley, 1988.

\bibitem{PhysRev.113.1593}
J.~O. Rasmussen, ``Alpha-decay barrier penetrabilities with an exponential
  nuclear potential: Even-even nuclei,'' {\em Phys. Rev.}, vol.~113,
  pp.~1593--1598, Mar 1959.

\bibitem{PhysRev.119.1069}
H.~J. Mang, ``Calculation of $\ensuremath{\alpha}$-transition probabilities,''
  {\em Phys. Rev.}, vol.~119, pp.~1069--1075, Aug 1960.

\bibitem{PhysRev.169.818}
K.~Harada and E.~A. Rauscher, ``Unified theory of alpha decay,'' {\em Phys.
  Rev.}, vol.~169, pp.~818--824, May 1968.

\bibitem{SANDULESCU1978205}
A.~Sandulescu, I.~Silisteanu, and R.~Wünsch, ``Alpha decay within feshbach
  theory of nuclear reactions,'' {\em Nuclear Physics A}, vol.~305, no.~1,
  pp.~205--212, 1978.

\bibitem{PhysRevLett.59.262}
S.~A. Gurvitz and G.~Kalbermann, ``Decay width and the shift of a
  quasistationary state,'' {\em Phys. Rev. Lett.}, vol.~59, pp.~262--265, Jul
  1987.

\bibitem{Holstein1996}
B.~R. Holstein, ``Understanding alpha decay,'' {\em American Journal of
  Physics}, vol.~64, pp.~1061--1071, 08 1996.

\bibitem{Kumar2025}
H.~Kumar and P.~Jain, ``Resonant nuclear fusion at second order,'' {\em Phys.
  Rev. C}, vol.~112, p.~014624, Jul 2025.

\bibitem{Miley1996NUCLEARTI}
G.~H. Miley and J.~A. Patterson, ``Nuclear transmutations in thin-film nickel
  coatings undergoing electrolysis,'' 1996.

\bibitem{SRINIVASAN2020233}
M.~Srinivasan and K.~Rajeev, ``Chapter 13 - transmutations and isotopic shifts
  in lenr experiments,'' in {\em Cold Fusion} (J.-P. Biberian, ed.), pp.~233 --
  262, Elsevier, 2020.

\bibitem{ramkumar2024low}
K.~Ramkumar, H.~Kumar, and P.~Jain, ``Low energy nuclear reactions through weak
  interactions,'' {\em arXiv preprint arXiv:2406.11550}, 2024.

\end{thebibliography}

\end{document}